\documentclass[10pt]{bmc_article}

\usepackage{cite} 
\usepackage{url}  
\usepackage{ifthen}  
\usepackage{multicol}   
\usepackage[utf8]{inputenc} 
\usepackage{mathtools}
 
\usepackage{graphicx}

\newboolean{publ}

\newenvironment{bmcformat}{\baselineskip20pt\sloppy\setboolean{publ}{false}}{\baselineskip20pt\sloppy}

\begin{document}
\begin{bmcformat}


\title{DGEclust: differential expression analysis of clustered count data}
 

\author{
		 Dimitrios V Vavoulis\correspondingauthor$^1$
         \email{Dimitris.Vavoulis@bristol.ac.uk}
       and 
         Margherita Francescatto$^2$
		 \email{Margherita.Francescatto@dzne.de}
       and 
	   	 Peter Heutink$^2$
         \email{Peter.Heutink@dzne.de}
       and 
         Julian Gough\correspondingauthor$^1$%
         \email{Julian.Gough@bristol.ac.uk}
}


\address{%
    \iid(1)Department of Computer Science, University of Bristol, Bristol, UK \\
	\iid(2)Genome Biology of Neurodegenerative Diseases, Deutsches Zentrum für Neurodegenerative Erkrankungen, Tübingen, Germany.
}%

\maketitle


\begin{abstract}
Most published studies on the statistical analysis of count data generated by 
next--generation sequencing technologies have paid surprisingly little attention 
on cluster analysis. We present a statistical methodology (\emph{DGEclust}) for 
clustering digital expression data, which (contrary to alternative methods) 
simultaneously addresses the problem of model selection (i.e. how many clusters 
are supported by the data) and uncertainty in parameter estimation. We show how 
this methodology can be utilised in differential expression analysis and we demonstrate 
its applicability on a more general class of problems and higher accuracy, when compared 
to popular alternatives. \emph{DGEclust} is freely available at 
\url{https://bitbucket.org/DimitrisVavoulis/dgeclust}
\end{abstract}


\ifthenelse{\boolean{publ}}{\begin{multicols}{2}}{}


\section*{Background}

Next--generation (NGS) or high--throughput sequencing (HTS) technologies provide a
revolutionary tool for the study of the genome, epigenome and transcriptome in a
multitude of organisms (including humans) by allowing the relatively rapid production
of millions of short sequence tags, which mirror particular aspects of the molecular state
of the biological system of interest. A common application of NGS technologies 
is the study of the transcriptome, which involves a family of methodologies, such as 
RNA-seq\cite{Wang:2009fk}, CAGE (Cap Analysis of Gene 
Expression; \cite{Shiraki:2003uq}), 
SAGE (Serial Analysis of Gene Expression; \cite{Velculescu:1995kx}) and others. 
Most published studies on the statistical analysis 
of count data generated by NGS
technologies have focused on the themes of experimental design\cite{Auer:2010uq},
normalisation\cite{Sun:2012kx,Dillies:2012vn} and the development of tests for differential
expression\cite{Oshlack:2010vn, Kvam:2012hc, Soneson:2013ys}. Surprisingly, not much
attention has been paid on cluster analysis. 

Clustering is considered an important tool in the study of genomic data and it has been 
used extensively in the analysis of microarrays\cite{Cho:1998ly,Eisen:1998ve,Shannon:2003qf} 
(see \cite{Jiang:2004zr} for a review of different clustering methods). It involves 
grouping together the expression profiles of different genes across different points in 
time, treatments and tissues, such that expression profiles in the same group are more 
similar in some way to each other than to members of other groups. Genes which are clustered 
together across samples exhibit co-related expression patterns, which might indicate 
co--regulation and involvement of these genes in the same cellular 
processes\cite{Yeung:2004bh}. Moreover, whole samples of gene expression profiles can 
be clustered together, indicating a particular macroscopic phenotype, such as 
cancer\cite{Souto:2008dq}.

A large class of clustering methods relies on the definition of a distance metric, which 
quantifies the ``similarity'' between any two gene expression data points. Subsequently, 
clusters are formed, such that the distance between any two data points in the same cluster 
is minimised. Typical methods in this category are K--means clustering and self--organising 
maps (SOMs) \cite{Jiang:2004zr}. Another important category includes model--based clustering 
algorithms. In this case, the whole gene expression dataset is modeled as a random sample 
from a finite mixture of probability distributions, where each component of the mixture 
corresponds to a distinct cluster. The parameters of each component in the mixture (e.g. 
mean and variance) are usually estimated using an Expectation--Maximization 
algorithm\cite{Jiang:2004zr}. Hierarchical clustering is yet a third type of clustering 
methodology, which is particularly suited for modelling genomic (often hierarchically 
organized) data. It generates a hierarchical series of nested clusters, which can be 
represented graphically as a \emph{dendrogram}. This stands in contrast to partition--based 
methods (e.g. K--means or SOMs), which decompose the data directly into a finite number 
of non--overlapping clusters\cite{Jiang:2004zr}.

In this paper, we present a model--based statistical approach and associated software
(\emph{DGEclust}) for clustering digital expression data. The proposed methodology
is novel, because it simultaneously addresses the problem of model selection
(i.e. how many clusters are supported by the data) and uncertainty (i.e. the error 
associated with estimating the number of clusters and the parameters of each cluster). 
This is possible by exploiting a Hierarchical Dirichlet Process Mixture Model or 
HDPMM\cite{Teh:2006dz}, a statistical framework, which has been applied in the past on 
multi--population haplotype inference\cite{Sohn:2009fv} and for modelling multiple text 
corpora\cite{Wang2011}. In our version of the HDPMM, individual expression profiles are 
drawn from the Negative Binomial distribution (as, for example, in
\cite{Anders:2010kl,Hardcastle:2010tg,McCarthy:2012ve}) and parameter estimation is
achieved using a novel blocked Gibbs sampler, which permits efficiently processing large 
datasets (including more than 20K features). We show how the output of our clustering 
algorithm can be utilised in differential expression analysis and, using simulated data,
we demonstrate its superior performance -- in terms of Receiver Operating characteristic 
(ROC) and False Discovery Rate (FDR) curves -- when compared to popular alternative methods.      
When applied on CAGE data from human brains, our methodology manages to detect a 
significantly larger number of differentially expressed transcripts than alternative
methods. An early version of the proposed methodology has been presented previously in 
poster format and in \cite{Vavoulis2013a}.      
 
\section*{Results and discussion}

\subsection*{Description of the model}

Formally, the production of count data using next--generation sequencing assays can be 
thought of as random sampling of an underlying population of cDNA fragments. 
Thus, the counts for each tag describing a class of cDNA fragments can, in principle, be 
modelled using the Poisson distribution, whose variance is, by definition, equal to its 
mean. However, it has been shown that, in real count data of gene expression, the variance 
can be larger than what is predicted by the Poisson distribution\cite{Lu:2005dq,Robinson:2007bh,Robinson:2008cr,Nagalakshmi:2008qf}. 
An approach that accounts for the so--called ``over-dispersion'' in the data is to adopt
quasi--likelihood methods, which augment the variance of the Poisson distribution with a 
scaling factor, thus dropping the assumption of equality between the mean and 
variance\cite{AueDoe11,Srivastava:2010nx,Wang:2010oq,Langmead:2010ly}. An alternative 
approach is to use the Negative Binomial distribution, which is derived from the Poisson, 
assuming a Gamma--distributed rate parameter. The Negative Binomial distribution 
incorporates both a mean and a variance parameter, thus modelling over--dispersion in 
a natural way\cite{Anders:2010kl,Hardcastle:2010tg,McCarthy:2012ve}. For this reason, 
in this paper we use the Negative Binomial distribution for modelling count data.

We indicate the number of reads for the $i^{th}$ feature (e.g. gene) at the $s^{th}$ 
sample/library of the $j^{th}$ class of samples (e.g. tissue or experimental condition) 
with the variable $y_{jsi}$. In a normalised dataset, we assume that $y_{jsi}$ is 
distributed according to a Negative Binomial distribution with gene-- and class--specific 
parameters $\theta_{ji}=\{\alpha_{ji}, p_{ji}\}$:
\begin{equation}
y_{jsi}|\theta_{ji} \sim
\frac{\Gamma(y_{jsi}+\alpha_{ji})}{\Gamma(\alpha_{ji})\Gamma(y_{jsi}+1)}
p_{ji}^{\alpha_{ji}}
\left(1 - p_{ji}\right)^{y_{jsi}}\label{eq:negbin}
\end{equation}
\noindent where $p_{ji} = \alpha_{ji} / \left(\alpha_{ji} + \mu_{ji}\right)$ is 
a probability measure, $\mu_{ji}$ is the (always positive) mean of the distribution and
$\alpha_{ji}$ is a dispersion parameter. Since, the variance 
$\sigma^2_{ji}=\mu_{ji}+\alpha_{ji}^{-1}\mu_{ji}^{2}$ is always larger 
than the mean by the quantity $\alpha_{ji}^{-1}\mu_{ji}^{2}$, the Negative Binomial 
distribution can be thought of as a generalisation of the Poisson distribution, which 
accounts for over--dispersion.   

\subsubsection*{Information sharing within samples}
A common limitation in experiments using NGS technologies is the low number or even 
absence of biological replicates, which complicates the statistical analysis of digital
expression data. One way to compensate for small sample sizes is to assume that all 
features share the same variance\cite{Robinson:2007bh}. A less restrictive approach is 
to implement some type of information sharing between features, which permits the improved 
estimation of the feature--specific over--dispersion parameters by pooling together features 
with similar expression profiles\cite{Anders:2010kl,Hardcastle:2010tg,McCarthy:2012ve}. 
In this paper, information sharing between features and between samples is introduced in a 
natural way due to the use of Dirichlet Processes as priors for the Negative Binomial 
distribution parameters and due to the hierarchical structure of the mixture model, as 
explained in this and the following section.
 
Specifically, within each sample class $j$, we assume that the set of gene--specific 
parameters $\{\theta_{ji}\}$ are random and distributed according to a prior 
distribution $G_j$, i.e.
\begin{equation}
\theta_{ji}|G_j\sim G_j
\label{eq:theta_prior}
\end{equation}
\noindent Furthermore, $G_j$ is itself randomly sampled from a {\it Dirichlet 
process} with positive {\it scaling parameter} $\gamma_j$ and {\it base probability} 
distribution $G_0$\cite{Teh:2006dz}:
\begin{equation}
G_j|\gamma_j,G_0 \sim \text{DP}(\gamma_j,G_0)
\label{eq:Gj}
\end{equation}
Dirichlet process priors are distributions over distributions and they have become a
popular choice in Bayesian inference studies, since they provide an elegant and, in
many ways, more realistic solution to the problem of determining the ``correct'' number
of components in mixture models. Standard theoretical results\cite{SETHURAMAN:1994aa}
state that a sample $G_j$ from Eq. \ref{eq:Gj} is a discrete distribution with probability
one over a countably infinite set of $\theta$s. Large values of $\gamma_j$ lead
to a large number of similarly likely values of $\theta$, while small
values of this parameter imply a small number of highly probable values of $\theta$. This
and Eq. \ref{eq:theta_prior} imply that the 
gene--specific parameters $\theta_{ji}$ are not all distinct. Different genes within the 
same class of libraries may share the same value of $\theta$ or, in other words, genes 
in class $j$ are grouped in a (not known in advance) number of clusters, based on the 
value of $\theta$ they share. Equivalently, the expression profiles of different groups 
of genes in a given class of samples are drawn from different Negative Binomial 
distributions, each characterised by its own unique value of $\theta$. This clustering 
effect within each sample class is illustarted in Fig. \ref{fig:overview}.   

\subsubsection*{Information sharing between samples} 

Up to this point, we have considered clustering of features (e.g. genes) within the same class of samples, 
but not across classes of samples (e.g. tissue or conditions). However, in a given dataset, 
each cluster might include gene expression profiles from the same, as well as from different 
sample classes. In other words, clusters are likely shared between samples that belong to 
different classes. This sharing of information between sample classes can be expressed 
naturally in the context of Hierarchical Dirichlet Process Mixture Models\cite{Teh:2006dz}. 
Following directly from the previous section, we assume that the base distribution 
$G_0$ is itself random and sampled from a Dirichlet Process with a global scaling parameter 
$\delta$ and a global base distribution $H$:
\begin{equation}
G_0|\delta,H \sim \text{DP}(\delta,H)
\label{eq:G0}
\end{equation}
\noindent This implies that $G_0$ is
(similarly to each $G_j$) discrete over a countably infinite set of atoms $\theta_k$, 
which are sampled from $H$, i.e. $\theta_k \sim H$. Since $G_0$ is the common base 
distribution of all $G_j$, the atoms $\theta_k$ are shared among all samples, yielding 
the desired information sharing across samples (see Fig. \ref{fig:overview}).                     

\subsubsection*{Generative model}
In summary, we have the following hierarchical model for the generation of a digital 
gene expression dataset (see also Fig. \ref{fig:overview}):
\begin{eqnarray}
G_0|\delta,H        & \sim & \text{DP}(\delta,H)	    		\nonumber \\
G_j|\gamma_j,G_0    & \sim & \text{DP}(\gamma_j,G_0)			\nonumber \\
\theta_{ji}         & \sim & G_j                        		\nonumber \\
y_{jsi}|\theta_{ji} & \sim & \text{NegBinomial}(\theta_{ji})
\label{eq:HDP}
\end{eqnarray}   
\noindent where the base distribution $H$ provides the global prior for sampling the 
atoms $\theta_k=(\alpha_k,p_k)$ and it takes the form of the following joint distribution:
\begin{equation}
\alpha^{-1}_k,p_k | \overbrace{\mu_\alpha, \sigma^2_\alpha}^\phi \sim 
\text{LogNormal}(\mu_\alpha,\sigma^2_\alpha) \cdot \text{Uniform}(0,1)	
\end{equation}
\noindent where $\phi$ is the set of hyperparameters, which $H$ depends on. According to 
the above formula, the inverse of the dispersion parameter $\alpha_k$ is sampled from a 
LogNormal distribution with mean $\mu_\alpha$ and variance $\sigma^2_\alpha$, while 
the probability parameter $p_k$ follows a Uniform distribution in the interval $[0,1]$. 
Given the above formulation, $\alpha_k$ is always positive, as it oughts to and, since the 
LogNormal distribution has a well known conjugate prior, the above particular form for $H$ 
greatly facilitates the posterior inference of the hyper-parameters $\phi$ (see below).
   
\subsection*{Inference}

The definition of the HDPMM in Eqs. \ref{eq:HDP} is implicit. In order to facilitate 
posterior inference, an equivalent constructive representation of the above model has been 
introduced in \cite{Wang2011} utilising Sethuraman's stick-breaking representation of a 
draw from a Dirichlet process\cite{SETHURAMAN:1994aa}. This representation introduces
a matrix of indicator variables $Z=\{z_{ji}\}$, where each element of the matrix, 
$z_{ji}$, indicates which cluster the $i^{th}$ expression measurement in the 
$j^{th}$ class of samples belongs to. Two different features belong to the same cluster 
if and only if their indicator variables, e.g. $z_{ji}$ and $z_{j'i'}$, are equal. 
A major aim of Bayesian inference in the above model, is to calculate the posterior 
distribution of matrix $Z$ given the dataset $Y$, i.e. $p(Z|Y)$. 

One approach to estimate this distribution is by utilizing Markov Chain Monte Carlo methods,
which generate a chain of random samples as a numerical approximation to the desired 
distribution. We have developed a blocked Gibbs sampler in the software package 
\emph{DGEclust}, which efficiently generates 
new samples from the posterior $p(Z|Y)$. The algorithm is an extension of 
the method presented in \cite{Ishwaran2001, Vavoulis2013a} for inference in non-hierarchical 
Dirichlet Process mixture models and its advantage is that it samples each element of $Z$ 
independently of all others. This not only results in very fast convergence, but it also
allows implementing the algorithm in vectorized form, which takes advantage 
of the parallel architecture of modern multicore processors and potentially permits 
application of the algorithm on very large datasets. Alternative MCMC methods, which are 
developed on the basis of the popular Chinese Restaurant Franchise representation of the 
HDP\cite{Teh:2006dz,Wang:2013ij}, do not enjoy the same advantage since they are restricted 
by the fact that sampling each indicator variable is conditioned on the remaining ones,
thus all of them must be updated in a serial fashion. Details of the algorithm are given in 
Vavoulis \& Gough, 2014 (in preparation).                             

\subsection*{Testing for differential expression}
Assuming that the above algorithm has been applied on a digital expression dataset $Y$ and
a sufficiently large chain of samples $Z^{(T_0+1)},Z^{(T_0+2)},\ldots,Z^{(T_0+T)}$ -- 
which approximates the posterior $p(Z|Y)$ -- has been generated, we show how these samples 
can be utilised in a differential expression analysis scenario. We consider two classes 
of samples, 1 and 2, which might represent, for example, two different tissues or 
experimental conditions. 

A particular feature (gene or transcript) is said to be \emph{not} differentially expressed,
if its expression measurements in classes 1 and 2 belong to the same cluster. In more formal
language, we say that the expected conditional probability $\pi_i=p(nDE_i|Y)$ that 
feature $i$ is not differentially expressed given data $Y$ is equal to the expected 
probability $p(z_{1i}=z_{2i}|Y)$ that the indicator variables of feature $i$ in sample 
classes 1 and 2 have the same value. This probability can be approximated as a simple 
average over the previously generated MCMC samples $\{Z^{T_0+t}\}_{t=1,\ldots,T}$: 
\begin{equation}
	\pi_i = \frac{\sum_{t=T_0+1}^{T_0+T} 1\!\!1\left(z_{1i}^{(t)}=z_{2i}^{(t)}\right)}{T}
\label{eq:pi}
\end{equation}
where $1\!\!1(\cdot)$ is equal to $1$ if the expression inside the parentheses is true 
and $0$ otherwise. Given a threshold $\tilde\pi$, we can generate a set $\mathcal{D}$ of 
potentially differentially expressed features with probabilities less than this threshold, 
i.e. $\mathcal{D}=\{i: \pi_i\le\tilde\pi\}$, where $\pi_i$ is calculated as in 
Eq. \ref{eq:pi} for all $i$.

As observed in \cite{Newton:2004aa}, the quantity $\pi_i$ measures the conditional 
probability that including the $i^{th}$ feature in list $\mathcal{D}$ is a Type I
error, i.e. a false discovery. This useful property makes possible the calculation of
the conditional False Discovery Rate (FDR) as follows:     
\begin{equation}
	FDR(\tilde\pi) = \frac{\sum_i \pi_i 1\!\!1\left(\pi_i\le\tilde\pi\right)}
	{\sum_i 1\!\!1\left(\pi_i\le\tilde\pi\right)}
\label{eq:FDR}
\end{equation}
From Eq. \ref{eq:FDR}, it can be seen that $\mathcal{D}$ always has an FDR at most equal to 
$\tilde\pi$. Alternatively, one can first set a target FDR, say tFDR, and then find the 
maximum possible value of $\tilde\pi$, such that $FDR(\tilde\pi) \le tFDR$. 

Notice that, unlike alternative approaches, which make use of gene--specific p--values, 
this methodology does not require any correction for multiple hypothesis testing,
such as the Benjamini–-Hochberg procedure.
Although the computation of FDR using Eq.~\ref{eq:FDR} is approximate (since it depends 
on the accuracy of the calculation of $\pi_i$ using Eq.~\ref{eq:pi}), it is reasonable to 
assume that the error associated with this approximation is minimised, if sufficient care 
is taken when postprocessing the MCMC samples generated by the Gibbs sampler. 

\subsection*{Application on clustered simulated data}
In order to assess the performance of our methodology, we applied it on simulated and
actual count data and we compared our results to those obtained by popular software 
packages, namely \emph{DESeq}, \emph{edgeR} and \emph{baySeq}.

First, we applied our algorithm on simulated clustered data, which were modelled after
RNA--seq data from yeast (\emph{Saccharomyces cerevisiae})
cultures\cite{Nagalakshmi:2008qf}. This data were generated using two different 
library preparation protocols. Samples for each protocol included two 
biological (i.e. from different cultures) and one technical (i.e. from the same culture) 
replicates, giving a total of six libraries. 

The simulated data were generated as follows: first, we fitted the yeast RNA--seq data 
with a Negative Binomial mixture model, where each component in the mixture was
characterised by its own $\alpha$ and $p$ parameters. We found that 10 
mixture components fitted the data sufficiently well. At this stage, it was not necessary
to take any replication information into account. The outcome of this 
fitting process was an optimal estimation of the parameters of each component in the 
mixture, $\alpha_k$ and $p_k$, and of the mixture proportions, $w_k$, where 
$k=1,\ldots,10$.

In a second stage, we generated simulated data using the fitted mixture model as 
a template. For each simulated dataset, we assumed 2 different classes of samples 
(i.e. experimental conditions or tissues) and 
2, 4 or 8 samples (i.e. biological replicates) per class. For gene $i$ in class $j$, we 
generated an indicator variable $z_{ji}$ taking values from $1$ to $10$ with 
probabilities $w_1$ to $w_{10}$. Subsequently, for gene $i$ in sample $s$ in 
class $j$, we sampled expression profile (i.e. counts) $y_{jsi}$ from a Negative 
Binomial distribution with parameters $\alpha_{z_{ji}}$ and $p_{z_{ji}}$. The 
process was repeated for all genes in all samples in both classes resulting in 
a matrix of simulated count data. Mimicking the actual data, the depth of each 
library was randomly selected between $1.7 \times 10^6$ to $3 \times 10^6$ reads.    

Gene $i$ was considered differentially expressed, if indicator variables
$z_{1i}$ and $z_{2i}$ were different, i.e. if the count data for gene $i$
in the two different classes belonged to different clusters. By further setting 
$z_{1i}$ equal to $z_{2i}$ for arbitrary values of $i$, it was possible to generate 
datasets with different proportions of differentially expressed genes. Since the 
proportion of differentially expressed genes may affect the ability of a method to
identify these genes\cite{Soneson:2013ys}, we examined datasets with 10\% and 50\% 
of their genes being differentially expressed.      

In our comparison of different methodologies, we computed the average Receiver 
Operating Characteristic (ROC) and False Discovery Rate (FDR) curves over a set 
of 10 different simulated datasets. In order to keep execution times to a reasonable 
minimum, we considered datasets with 1000 features. All examined methods allow to rank 
each gene by providing nominal p--values 
(\textit{edgeR}, \textit{DESeq}) or posterior probabilities (\textit{DGEclust}, 
\textit{baySeq}). Given a threshold score, genes on opposite sides of the threshold 
are tagged as differentially expressed or non--differentially expressed, accordingly. 
In an artificial dataset, the genes that were simulated to be differentially expressed 
are considered to be the true positive group, while the remaining genes are considered 
the true negative group. By computing the False Positive Rate (FPR) and the 
True Positive Rate (TPR) for all possible score thresholds, we can construct ROC and 
FDR curves for each examined method. The area under a ROC curve is a measure of the 
overall discriminative ability of a method (i.e. its ability to correctly classify 
transcripts as differentially or non--differentially expressed). Similarly, the area 
under an FDR curve is inversely related to the discriminatory performance of 
a classification method.  

Our results are summarised in Fig.~\ref{fig:clustered}. When datasets with 10\% 
DE transcripts and 
a small number of samples is considered (Fig.~\ref{fig:clustered}Ai), 
\textit{DGEclust} performs 
better than the similarly performing \textit{baySeq}, \textit{edgeR} and \textit{DESeq}. 
By increasing the number of samples to 4 and 8 (Figs.~\ref{fig:clustered}Aii and Aiii), 
we can increase
the discriminatory ability of all methods. Again, \textit{DGEclust} is the top performer, 
with \textit{baySeq} following closely.  

While ROC curves provide an overall measure of the discriminatory ability of differrent 
classification methods, they do not immediately indicate whether the deviation 
from a perfect classification is mainly due to false positives or false negatives. 
For this reason, we also constructed FDR curves, which illustrate the number of false 
positives as a function of the total number of positives (i.e. as the decision threshold 
changes). Mean FDR curves for datasets with 10\% DE transcripts are illustrated 
in Figs.~\ref{fig:clustered}Bi--Biii. Notice that we measure the false positives only 
among the 
first 100 discoveries, which is the true number of DE transcripts in the simulated
datasets. We may observe that \textit{DGEclust} keeps the number of false positives 
smaller than the corresponding number of the examined competitive methods. This is 
particularly true at a small number of samples (Figs.~\ref{fig:clustered}Bi and Bii). 
For a large number
of samples (Fig.~\ref{fig:clustered}Biii), \textit{DGEclust} and \textit{baySeq} 
perform similarly in 
terms of keeping false positives to a minimum among the top ranking transcripts.

The same trends are observed when we considered datasets with 50\% DE transcripts. 
In this case, the difference in performance between \textit{DGEclust} and the competitive
methods is even more prominent, as indicated by the average ROC curves 
(Figs.~\ref{fig:clustered}Ci--Ciii). This is mainly due to a drop in the performance 
of \textit{DESeq}, 
\textit{baySeq} and \textit{edgeR} and not to an increase in the performance of 
\textit{DGEclust}, which remains largely unaffected. This is particularly true when a larger 
number of samples is considered (Figs.~\ref{fig:clustered}Cii,Ciii). In terms of keeping 
the false positives to a minimum among the top--ranking trascripts, \textit{DGEclust} is 
again the top performer, with \textit{baySeq} in the second place
(Figs.~\ref{fig:clustered}D). 
Notice that when a large number of samples is available, \textit{DGEclust} does not return 
any false positives among the first 500 discoveries (Fig.~\ref{fig:clustered}Diii), 
which is the true number of DE transcripts in the simulated datasets.     

\subsection*{Application on unclustered simulated data}

Since our algorithm is designed to take advantage of the cluster structure that may 
exist in the data, testing different methods on clustered simulated data might give
an unfair advantage to \textit{DGEclust}. For this reason, we further tested the above 
methodologies on unclustered simulated data (or, equivalently, on simulated data,
where each gene is its own cluster). As in the case of clustered simulated data, 
the unclustered data were modelled after yeast RNA--seq data\cite{Nagalakshmi:2008qf}, 
following a procedure similar to \cite{Soneson:2013ys}. In a first stage, we used 
\emph{DESeq} to estimate unique $\alpha_i$ and $p_i$ parameters for each gene 
$i$ in the yeast data. In a second stage, for each gene $i$ in each class $j$ in 
the simulated data, we sampled randomly a unique index $z_{ji}$ ranging from $1$ to 
$N$, where $N$ was the total number of genes in the simulated data. Subsequently, 
for each gene $i$ in each sample $s$ in each class $j$, we sampled counts $y_{jsi}$ 
from a Negative Binomial distribution with parameters $\alpha_{z_{ji}}$ and 
$p_{z_{ji}}$. As for the clustered data, gene $i$ was simulated to be differentially 
expressed by making sure that the randomly sampled indicator variables $z_{i1}$ and 
$z_{i2}$ had different values. The above procedure for simulating unclustered count 
data makes minimal assumptions about the expression profiles of differentially 
and non--differentially expressed genes and it is a simple extension of the procedure 
we adopted for simulating clustered count data. As above, we considered datasets 
with 1000 genes, 2 sample classes and 2, 4 or 8 samples per class and we randomly selected
the library depths between $1.7 \times 10^6$ and $3 \times 10^6$ reads. Also, datasets 
with either 10\% or 50\% of their genes being differentially expressed were considered.

Our results from testing \textit{DGEcust}, \textit{edgeR}, \textit{DESeq} and 
\textit{baySeq} on unclustered simulated data are presented in Fig.~\ref{fig:unclustered}.
We may observe that all methods perform similarly for both low 
(Fig.~\ref{fig:unclustered}A,B) and high (Fig.~\ref{fig:unclustered}C,D) proportions of 
DE genes. In particular, despite the absence of a clear 
cluster structure in the data, \textit{DGEclust} is not worse than competitive methods. 
This is indicative of the fact that our algorithm is applicable on a more general 
category of count datasets, which includes either clustered or unclustered data. 
As in the case of clustered data, increasing the number of samples improves the overall 
performance of the various methodologies (Figs.~\ref{fig:unclustered}A) and reduces 
the number of false positives among the top--ranking discoveries 
(Figs.~\ref{fig:unclustered}B). The same trends are observed, when a high proportion of 
DE genes is considered (Figs.~\ref{fig:unclustered}C,D).   

\subsection*{Application on CAGE human data}
In addition to simulated data, we also tested our method on CAGE libraries, which
were prepared according to the standard Illumina protocol described in \cite{Takahashi:2012aa}.
The dataset consisted of twenty--five libraries isolated from five brain regions 
(caudate nucleus, frontal lobe, hippocampus, putamen and temporal lobe) from five human
donors and it included 23448 features, i.e. tag clusters representing promoter regions
(see Materials and Methods for more details). 

As illustrated in Fig.~\ref{fig:sim_progress}A, \textit{DGEclust} was left to process 
the data for 10K iterations. Equilibrium was attained rapidly, with the estimated number 
of clusters fluctuating around a mean value of approximatelly 81 clusters 
(Figs.~\ref{fig:sim_progress}A,B)
and the normalised autocorrelation of the Markov chain dropping quickly below 0.1 after 
a lag of only around 10 iterations (Fig.~\ref{fig:sim_progress}C). A snapshot of the 
fitted model at the end of the 
simulation illustrates how samples from each brain region are tightly approximated 
as mixtures of Negative Binomial distributions (i.e. clusters; 
Fig.~\ref{fig:fitted}). After
the end of the simulation, we used the procedure outlined in a previous section in order
to identify differentially expressed transcripts using a nominal burn--in period of 
$T_0=1000$ iterations. We also applied \textit{edgeR}, \textit{DESeq} and 
\textit{baySeq} on the same
dataset in order to find differentially expressed transcripts between all possible pairs 
of brain regions. Transcripts selected at an FDR threshold of 0.01 were considered 
differentially expressed. 

In a first stage, we compared the number of DE transcripts 
found by different methods. It can be observed (Fig.~\ref{fig:matrix_plot}, upper triangle) 
that, for all 
pairs of brain regions, \textit{DGEclust} returned the largest number of DE transcripts, 
followed by \textit{edgeR} and, then, \textit{DESeq} and \textit{baySeq} which, for all 
cases, discovered a similar number of DE transcripts. In particular, \textit{DGEclust} was 
the only method that found a significant number of DE transcripts (520) between the 
frontal and temporal lobes, whereas \textit{edgeR}, \textit{DESeq} and \textit{baySeq} 
found only 7, 3 and 4 DE genes, respectively. By checking the overlap between transcripts
classified as DE by different methods (Fig.~\ref{fig:matrix_plot}, lower triangle), 
we conclude that the DE 
transcripts found by \textit{edgeR}, \textit{DESeq} and \textit{baySeq} are in all
cases essentially a subset of the DE transcripts discovered by \textit{DGEclust}. 
\textit{DGEclust} appears to identify a large number of ``unique'' transcripts, in addition 
to those discovered by other methods, followed in this respect by \textit{edgeR}, which 
also found a small number of ``unique'' transcripts in each case. 

In a second stage, we compared the number of DE genes identified by \textit{DGEclust} 
between different brain regions and we constructed a (symmetric) similarity matrix, 
which can be used as input to hierarchical clustering routines for the generation of 
dendrograms and heatmaps. Specifically, each element $s_{j_1j_2}$ of this matrix  
measuring the similarity between brain regions $j_1$ and $j_2$ is defined as follows: 
\begin{equation}
	s_{j_1j_2} = \frac{\sum_i\pi_i}{N}\biggr|_{j_1j_2}
\label{eq:similarity}
\end{equation}
\noindent where $N$ is the number of features in the dataset and $\pi_i$ is the 
probability that transcript $i$ is differentially expressed between regions $j_1$ and
$j_2$, as computed by Eq.~\ref{eq:pi}. The similarity matrix calculated as above was used to 
construct the dendrogram/heatmap in Fig.~\ref{fig:hier}, after employing a cosine distance 
metric and complete linkage. 

It may be observed that the resulting hierarchical clustering reflects the
evolutionary relations between different regions. For example, the temporal and 
frontal lobe samples, which are both located in the cerebral 
cortex, are clustered together and they are maximally distant from subcortical regions, 
such as the hippocampus, caudate nucleus and putamen. The last two are also clustered 
together and they form the dorsal striatum of the basal ganglia.

\section*{Conclusions}
Despite the availability of several protocols (e.g. single vs. paired--end)
and sequencing equipment (e.g. Solexa's Illumina Genome Analyzer, ABI Solid Sequencing 
by Life Technologies and Roche's 454 Sequencing), all NGS technologies follow a common 
set of experimental steps (see \cite{Oshlack:2010vn} for a review) and, eventually,
generate data, which essentially constitutes a discrete, or \emph{digital} measure of 
gene expression. This data is fundamentally different in nature (and, in general terms, 
superior in quality) from the continuous fluorescence intensity measurements obtained 
from the application of microarray technologies. In comparison, NGS methods
offer several advantages, including detection of a wider level of expression
levels and independence on prior knowledge of the biological system, which is required by
the hybridisation--based microarrays\cite{Oshlack:2010vn}. Due to 
their better quality, next--generation sequence assays tend to replace microarrays, 
despite their higher cost\cite{Carninci:2009fk}. 

In this paper, we have addressed the important issue of clustering digital expression data, 
a subject which is surprisingly lacking in methodological approaches, when compared to 
micro--arrays. Most proposals for clustering RNA--seq and similar types of data have 
focused on clustering 
variables (i.e. biological samples), instead of features (e.g. genes) and they employ 
distance--based or hierarchical clustering methodologies on appropriatelly transformed 
datasets, e.g. \cite{Anders:2010kl,Severin:2010cr,Li:2010nx}. For example, the authors 
in \cite{Anders:2010kl} calculate a common variance function for all samples in a 
Tag--seq dataset of glioblastoma--derived and non--cancerous neural stem cells using a 
variance--stabilizing transformation, followed by hierarchical clustering using a Euclidean 
distance matrix. In \cite{Severin:2010cr}, a Pearson correlation dissimilarity metric 
was used for the hierarchical clustering of RNA--seq profiles in 14 different tissues of 
soybean after these were normalised using a variation of the RPKM 
method\cite{Sun:2012kx,Dillies:2012vn}. 

The above approaches, although fast and relatively easy to implement, do not always take 
into account the discrete nature of digital gene expression data. For this reason, various 
authors have developed distance metrics based on different parameterizations 
of the log--linear Poisson model for modelling count data, 
e.g. \cite{Cai:2004kl,Kim:2007oq,Witten:2011tg}. A more recent class of methods 
follows a model--based approach, where the digital dataset is modeled as a random sample 
from a finite mixture of discrete probability distributions, usually Poisson or 
Negative Binomial\cite{Si2011,Rau2011,Wang:2013hc}. Utilising a full statistical 
framework for describing the observed count data, these model--based approaches 
often perform better than distance--based algorithms, such as K-means\cite{Si2011}.

Although computationally efficient and attractive due to their relative conceptual 
simplicity, the utility of both distance-- and finite model--based clustering methods 
has been critisised\cite{Rasmussen:2009bs,Wang:2013ij}. One particular feature of 
these methodologies, which compromises their applicability, is that the number of 
clusters in the data must be known \emph{a priori}. For example, both the K--means 
and the SOM alogrithms require the number of clusters as input. Similarly, methods 
which model the data as a finite mixture of Poisson or Negative Binomial 
distributions\cite{Si2011,Rau2011,Wang:2013hc} require prior knowledge of
the number of mixture components. Estimating the number of clusters usually makes 
use of an optimality criterion, such as the Bayesian Information Criterion (BIC)
or the Akaike Information Criterion (AIC), which requires repeated application of 
the algorithm on the same dataset with different initial choices of the number of 
clusters. Thus, the number of clusters and the parameters for each individual cluster 
are estimated separately, making the algorithm sensitive to the initial model choice. 
Similarly, hierarchical clustering methods often rely on some arbitrary distance metric 
(e.g. Euclidian or Pearson correlation) to distinguish between members of different 
clusters, without providing a criterion for choosing the ``correct'' number of clusters 
or a measure of the uncertainty of a particular clustering, which would serve to 
assess its quality.

In this paper, we have developed a statistical methodology and associated 
software (\emph{DGEclust}) for clustering digital gene expression data, which 
(unlike previously published approaches
\cite{Severin:2010cr,Li:2010nx,Cai:2004kl,Kim:2007oq,Witten:2011tg})
does not require any prior knowledge on the number of clusters, 
but it rather estimates this parameter and its uncertainty simultaneously with the 
parameters (e.g. location and shape) of each individual cluster. This is achieved 
by embedding the Negative Binomial distribution for modelling count data in a 
Hierarchical Dirichlet Process Mixtures framework. This formulation implies that 
individual expression measurements in the same sample or in different samples may 
be drawn from the same distribution, i.e. they can be clustered together. This is a 
form of information sharing within and between samples, which is made
possible by the particular hierarchical structure of the proposed model. At each level 
of the hierachy, the number of mixture components, i.e. the number of clusters, 
is assumed infinite. This representes a substantial departure from previously proposed 
finite mixture models and avoids the need for arbitrary prior choices regarding the 
number of clusters in the data. 

Despite the infinite dimension of the mixture model, only the finite number of clusters 
supported by the data and the associated parameters are estimated.
This is achieved by introducing a blocked Gibbs sampler, which permits 
efficiently processing large datasets, containing more than 10K 
genes. Unlike MCMC inference methods for HDPMM based on the popular Chinese Restaurant 
Franchise 
metaphor\cite{Teh:2006dz}, our algorithm permits updating all gene-specific parameters in 
each sample simultaneously and independently from other samples. This allows rapid 
convergence of the algorithm and permits developing parallelised implementations of 
the Gibbs sampler, which enjoy the increased performance offered by modern multicore 
processors.

Subsequently, we show how the fitted model can be utilised in a differential expression
analysis scenario. Through comparison to popular alternatives on both simulated and actual
experimental data, we demonstrate the applicability of our method on both clustered and
unclustered data and its superior performance in the former case. In addition, we show
how our approach can be utilised for constructing library--similarity matrices, which can
be used as input to hierarchical clustering routines. A slight modification of 
Eq.~\ref{eq:similarity} can be used for constructing gene--similarity matrices 
(see Vavoulis \& Gough 2014, in preparation). Thus, our methodology can be used 
to perform both gene-- and sample--wise hierarchical clustering, in contrast to 
existing approaches, which are appropriate for clustering either 
samples\cite{Severin:2010cr,Li:2010nx} or genes\cite{Cai:2004kl,Kim:2007oq,Witten:2011tg} 
only.

In conclusion, we have developed a hierarchical, non--parametric Bayesian clustering 
method for digital expression data. The novelty of the method is simultaneously addressing
the problems of model selection and estimation uncertainty and it can be utilised in
testing for differential expression and for sample--wise (and gene--wise) 
hierarchical grouping. We expect our work to inspire and support further theoretical 
research on modelling digital expression data and we believe  
that our software, \emph{DGEclust}, will prove to be a useful addition to the existing 
tools for the statistical analysis of RNA--seq and similar types of data.          
  
\section*{Materials and methods}
We implemented the methodology presented in this paper as the software package 
\emph{DGEclust}, which is written in Python and uses the SciPy stack. \emph{DGEclust}
consists of three command--line tools: \emph{clust}, which expects as input and clusters 
a matrix of unnormalised count data along with replication information, if this is 
available; \emph{pvals}, which takes as input the output of
\emph{clust} and returns a ranking of features, based on their posterior probabilities 
of being differential expressed; \emph{simmat}, which also takes as input the output of 
\emph{clust} and generates a feature-- or library--wise similarity matrix, which can 
be used as input to hierarchical clustering routines for the generation of heatmaps and 
dendrograms. All three programs take advantage of multi--core processors in order to 
accelerate computations. Typical calculations take from a few minutes (as for the 
simulated data used in this study) to several hours (as for the CAGE data), depending
on the size of the dataset and total number of simulation iterations. All analyses in this 
paper were
performed using \emph{DGEclust} and standard Python/SciPy tools, as well as \emph{DESeq},
\emph{edgeR} and \emph{baySeq} for comparison purposes. 

\subsection*{URL}
The most recent version of \emph{DGEclust} is freely available at the following location:
\url{https://bitbucket.org/DimitrisVavoulis/dgeclust}
          
\subsection*{Normalisation}
\emph{DGEclust} uses internally the same normalisation method as \emph{DESeq},
unless a vector of normalisation factors is provided at the command 
line. When comparing different software packages, we used the default normalisation 
method in each package or the method provided by \emph{DESeq}, whichever permitted the 
best performance for the corresponding method.      

\subsection*{CAGE libraries preparation and data pre--processing}

Human post--mortem brain tissue from frontal, temporal, hippocampus,
caudate and putamen from 5 donors was obtained from the Netherlands
Brain Bank (NBB, Amsterdam, Netherlands). Total RNA was extracted and
purified using the Trizol tissue kit according to the manufacturer
instructions (Invitrogen).

CAGE libraries were prepared according to the standard Illumina CAGE
protocol\cite{Takahashi:2012aa}. Briefly, five micrograms of total
RNA was reverse transcribed with Reverse Transcriptase. Samples were
cap--trapped and a specific linker, containing a 3--bp recognition site
and the type III restriction--modification enzyme EcoP15I, was ligated
to the single--strand cDNA. The priming of the second strand was made
with specific primers. After second strand synthesis and cleavage with
EcoP15I, another linker was ligated. Purified cDNA was then amplified
with 10 to 12 PCR cycles. PCR products were purified, concentration
was adjusted to 10 nM and sequenced on the HiSeq 2000 using the
standard protocol for 50bp single end runs.
 
Sequenced reads (tags) were filtered for known CAGE artifacts using
TagDust\cite{Lassmann:2009aa}. Low quality reads and reads mapping to
known rRNA were also removed. The remaining reads were mapped to the human
genome (build hg19) using the Burrows--Wheeler Aligner for short reads\cite{Li:2009aa}. 
Mapped reads overlapping or located within 20 bp on the same strand were grouped 
into tag clusters and tag clusters with low read counts were removed.

\bigskip

\section*{Authors' contributions}
	DVV and JG developed the method. MF and PH provided the CAGE data. DVV implemented 
	the method. DVV and MF performed the analyses. DVV and JG wrote the paper with 
	contributions from all authors. The final manuscript was read and approved by 
	all authors.

\section*{Acknowledgements}
  \ifthenelse{\boolean{publ}}{\small}{}
  DVV would like to thank Peter Green, Mark Beaumont and Colin Campbell from the
  University of Bristol for useful discussions on the subjects of Dirichlet Process 
  Mixture Models and multiple hypothesis testing. 
 

\newpage
{\ifthenelse{\boolean{publ}}{\footnotesize}{\small}
 \bibliographystyle{bmc_article}  
  \bibliography{bmc_article} }     


\ifthenelse{\boolean{publ}}{\end{multicols}}{}



\section*{Figure legends}

\begin{figure}[h]
\caption{
\textbf{Information sharing within and between sample classes.} 
Within each sample class, the count expression profiles for each feature (e.g. gene) 
across all samples (replicates) follow a Negative Binomial distribution. Different genes 
within the same class share the same distribution parameters, which are randomly sampled 
from discrete, class--specific random distributions ($G_1$ and $G_2$ in the figure). 
This imposes a clustering effect on genes in each sample class; genes in the same
cluster have the same color in the figure, while the probability of each cluster is
proportional to the length of the vertical lines in distributions $G_1$ and $G_2$. 
The discreteness of $G_1$ and $G_2$ stems from the fact that they are random samples
themselves from a Dirichlet Process with global base distribution $G_0$, which is also
discrete. Since $G_0$ is shared among all sample classes, the clustering effect extends 
between classes, i.e. a particular cluster may include genes from the same and/or different
sample classes. Finally, $G_0$ is discrete, because it too is sampled from a Dirichlet 
Process with base distribution $H$, similarly to $G_1$ and $G_2$. If the expression 
profiles of a particular gene belong to two different clusters across two classes, 
then this gene is considered \emph{differentially expressed} (see rows marked with stars 
in the figure).      
}
\label{fig:overview}
\end{figure}

\begin{figure}[h]
\caption{
\textbf{Comparison of different methods on clustered simulated data.} 
We examined datasets where 10\% (A,B) or 50\% (C,D) of the transcripts were 
differentially expressed.
The Receiver Operating Characteristic (ROC; Ai--Aiii, Ci--Ciii) and False Discovery Rate 
(FDR; Bi--Biii, Di--Diii) curves are averages over 10 distinct simulated datasets. The 
dashed lines in figures Ai--Aiii and Ci--Ciii indicate the performance of a completely 
random classifier. In all cases where 10\% of the transcripts were differentially expressed
(A,B), \emph{DGEclust} was 
the best method, followed closely by \emph{baySeq}. 
\emph{edgeR} and \emph{baySeq} 
perform similarly to each other and occupy the third place. The discriminatory ability 
of all methods increases with the available number of samples. 
In datasets where 50\% of the transcripts were differentially expressed (C,D), 
\emph{DGEclust} shows the best 
discriminatory ability, followed by \emph{baySeq} and \emph{edgeR} 
in the second place and \emph{DESeq} in the third place, similarly to A and B. 
The larger proportion
of differentially expressed transcripts results in worse perofrmance for all methods, 
with the exception of \emph{DGEclust} (compare to Ai-Aiii and Bi-Biii). Notice that 
when 8 samples are available, \emph{DGEclust} does not return any false positives 
among the first 500 discoveries, which is the true number of differentially expressed 
transcripts in the datasets (Diii).
}
\label{fig:clustered}
\end{figure}

\begin{figure}[h]
\caption{
\textbf{Comparison of different methods on unclustered simulated data.}
As in Fig.~\ref{fig:clustered}, datasets with 10\% or 50\% differentially expressed 
transcripts were examined. Average ROC (A,C) and FDR (B,D) curves and dashed lines in A and
C are as in Fig.~\ref{fig:clustered}. All methods, including \emph{DGEclust}, perform 
similarly and their overall performance improves as an increasing number of samples 
(2, 4 and 8) is considered. This is true regardless of the proportion of the 
differentially expressed transcripts in the data.  
}
\label{fig:unclustered}
\end{figure}

\begin{figure}[h]
\caption{
\textbf{Number of clusters in the macaque CAGE data estimated by \emph{DGEclust}.}
The Markov chain generated by the Gibbs sampler converged rapidly (after 
less than 500 iterations) and remained stable around a mean value of 81 clusters
until the end of the simulation after 10K simulation steps (A). 
The histogram constructed from the random samples 
in A provides an approximation of the posterior probability mass function of the
number of clusters in the macaque CAGE data (B). We may observe that the data supports 
between 75 and 90 clusters.
The auto-correlation coefficient of the Markov chain in A drops rapidly below 
a value of 0.1 at a lag of only around 10 iterations (C).   	        
}
\label{fig:sim_progress}
\end{figure}

\begin{figure}[h]
\caption{
\textbf{A snapshot of the fitted model after 10K simulation steps.}
Each panel illustrates the histogram of the log--transformed counts of a random sample 
from each brain region. Each sample includes 23448 features (i.e. tag clusters 
corresponding to promoter regions) and it is modelled as a mixture of Negative 
Binomial distributions (i.e. clusters; the solid black lines in each panel). The 
overall fitted model for each sample (the red line in each panel) is the weighted 
sum of the Negative Binomial components estimated for this sample by \emph{DGEclust}.    
}
\label{fig:fitted}
\end{figure}

\begin{figure}[h]
\caption{
\textbf{Comparison of differentially expressed transcripts in the macaque CAGE data 
discovered by different methods.}
The number of differentially expressed transcripts discovered by different methods 
for each pair of brain regions is illustrated in the upper triangle of the plot matrix,
while the overlap of these discoveries is given in the lower triangle. For all pairs 
of brain regions, \emph{DGEclust} finds the largest number of differentially expressed 
transcripts, followed by \emph{edgeR} (upper triangle of the plot matrix). In addition, 
the set of all differentially expressed transcripts discovered by \emph{edgeR}, 
\emph{DESeq} and \emph{baySeq} is essentially a subset of those discovered by 
\emph{DGEclust} (lower triangle of the plot matrix). Among all methods, 
\emph{DGEclust} finds the largest number of ``unique'' DE genes, followed by 
\emph{edgeR}.           
}
\label{fig:matrix_plot}
\end{figure}

\begin{figure}[h]
\caption{
	\textbf{Hierarchical clustering of brain regions in the macaque CAGE data.}
	We constructed a similarity matrix based on the number of differentially
	expressed transcripts discovered by \emph{DGEclust} between all possible pairs 
	of brain regions. This similarity matrix was then used as input to a 
	hierarchical clustering routine using cosine similarity as 
	a distance metric and a complete linkage criterion resulting in the illustrated 
	heatmap and dendrograms. Cortical regions (frontal and temporal lobe) are clustered 
	together and are maximally distant from subcortical regions, i.e. the hipocampus
	and the dorsal striatum (putamen and caudate nucleus) of the basal ganglia.           
}
\label{fig:hier}
\end{figure}

\newpage


%
%
%
%
%

\begin{figure}[h]
\begin{center}
	\includegraphics{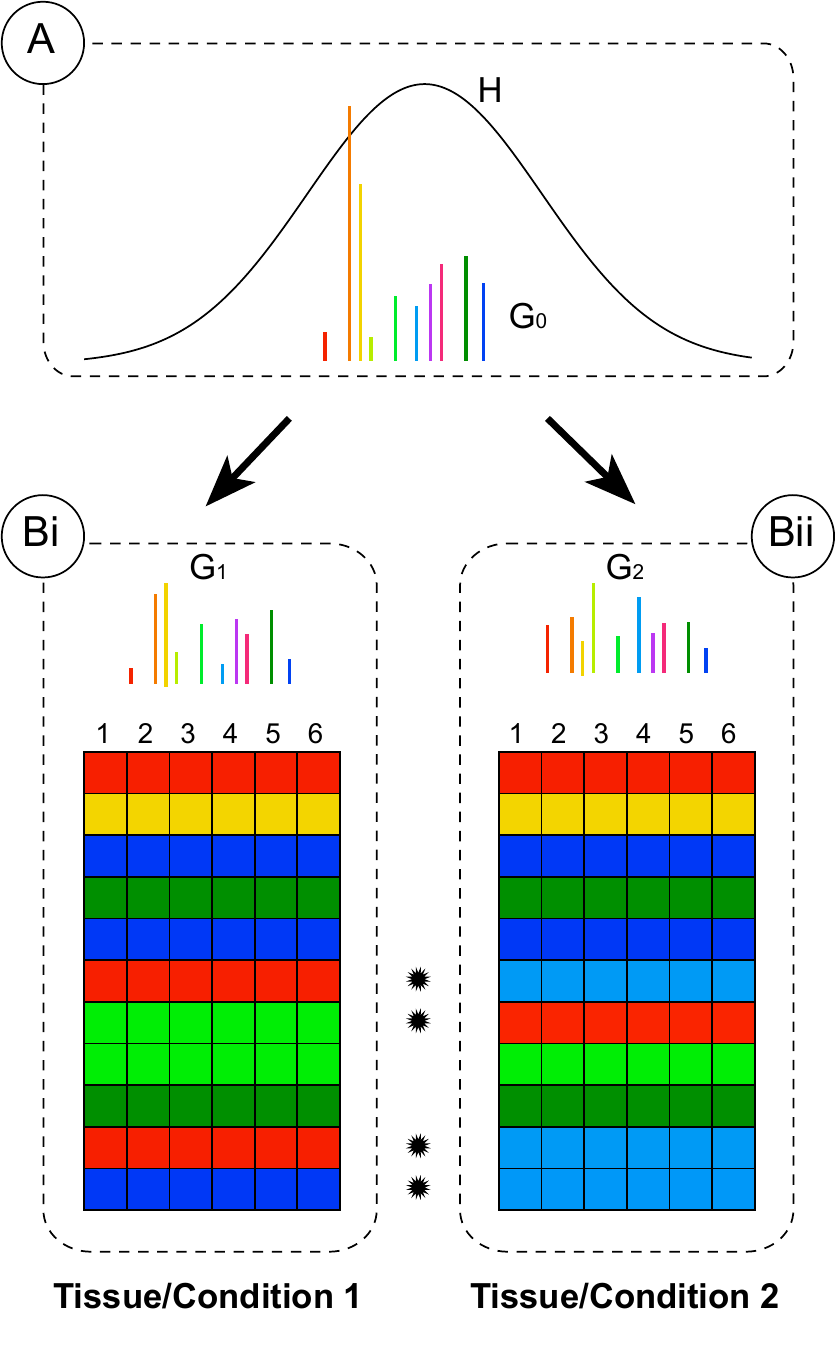} \\
	\textbf{Figure 1}
\end{center}
\end{figure}

\begin{figure}[h]
\begin{center}
	\includegraphics[width=0.95\textwidth]{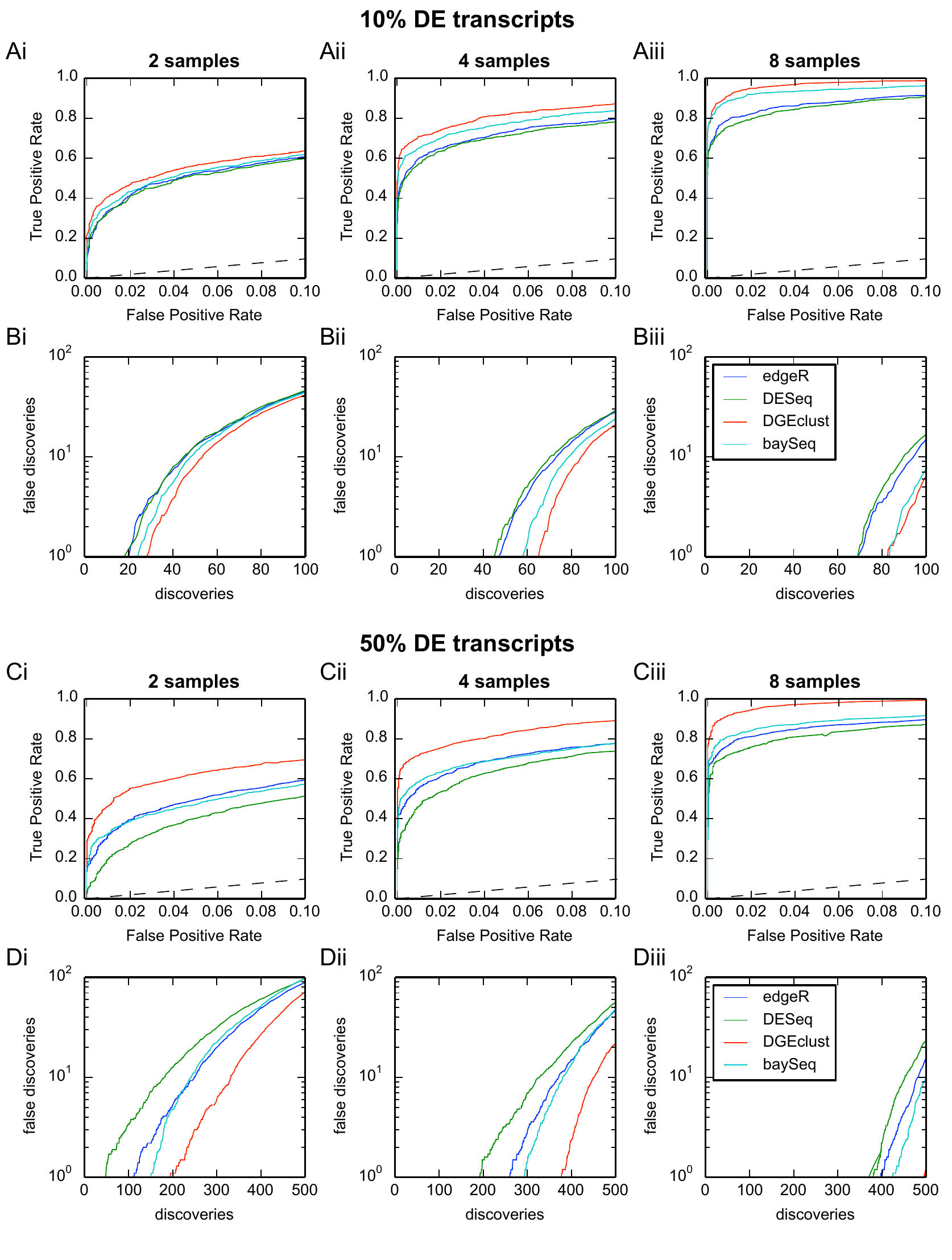} \\
	\textbf{Figure 2}
\end{center}
\end{figure}

\begin{figure}[h]
\begin{center}
	\includegraphics[width=0.95\textwidth]{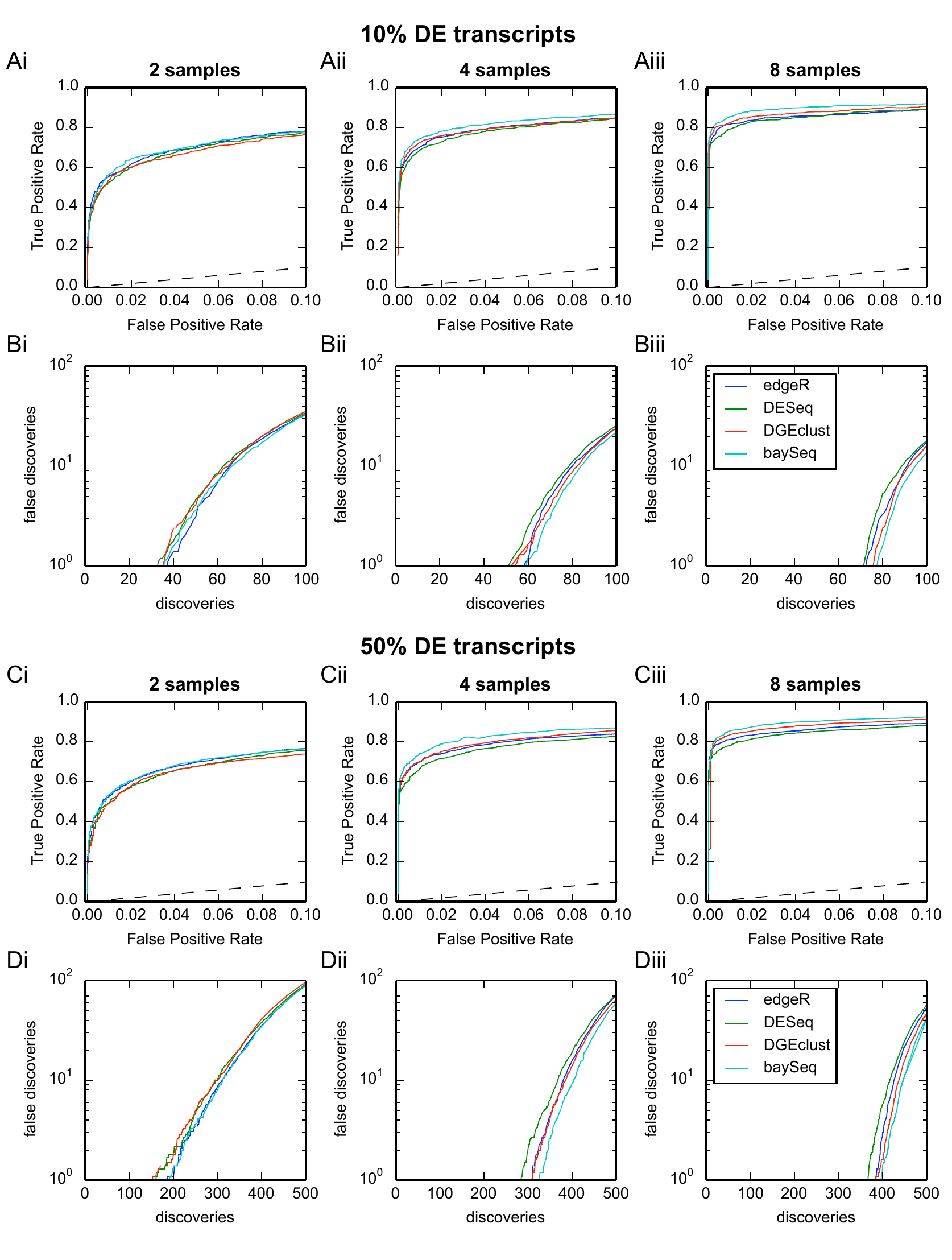} \\
	\textbf{Figure 3}
\end{center}	
\end{figure}

\begin{figure}[h]
\begin{center}
	\includegraphics{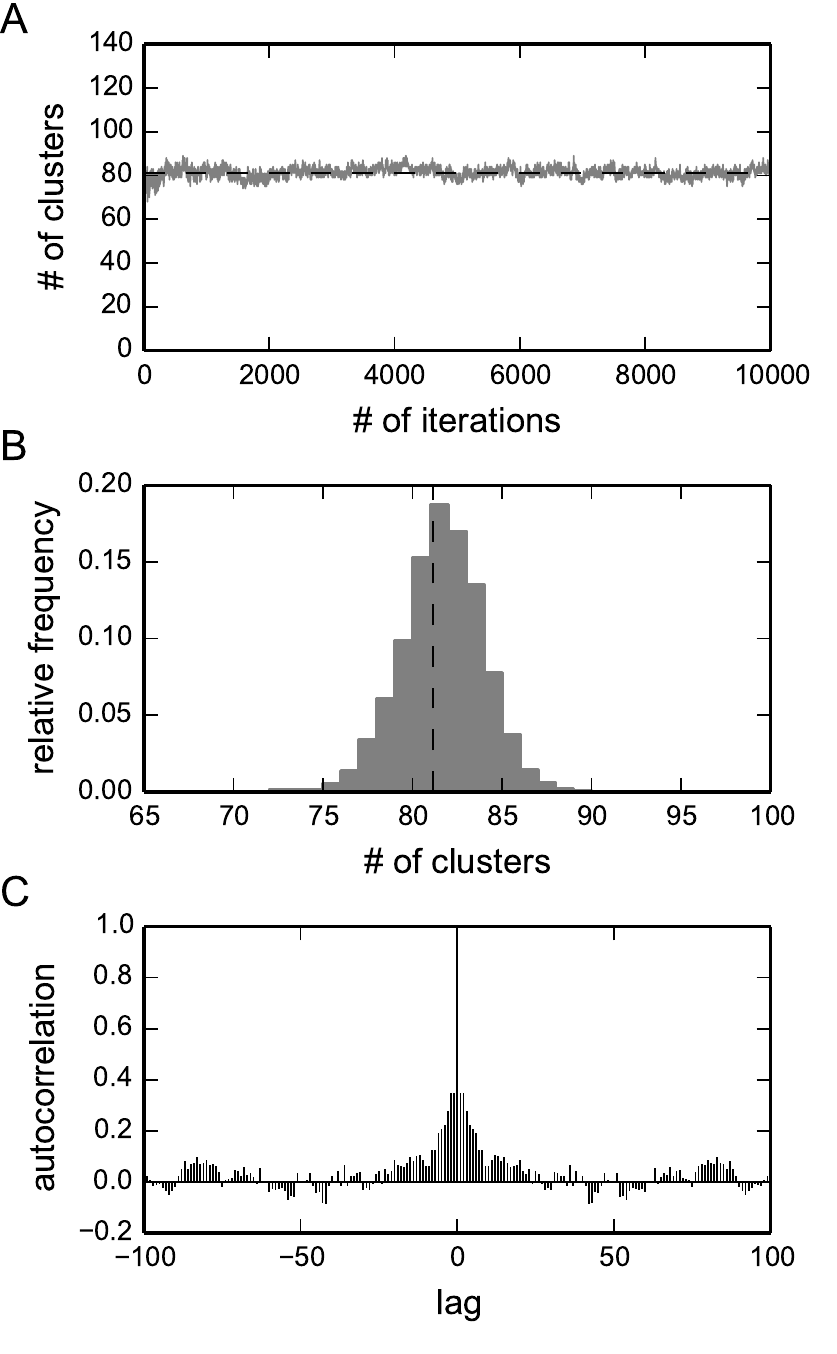} \\
	\textbf{Figure 4}
\end{center}	
\end{figure}

\begin{figure}[h]
\begin{center}
	\includegraphics{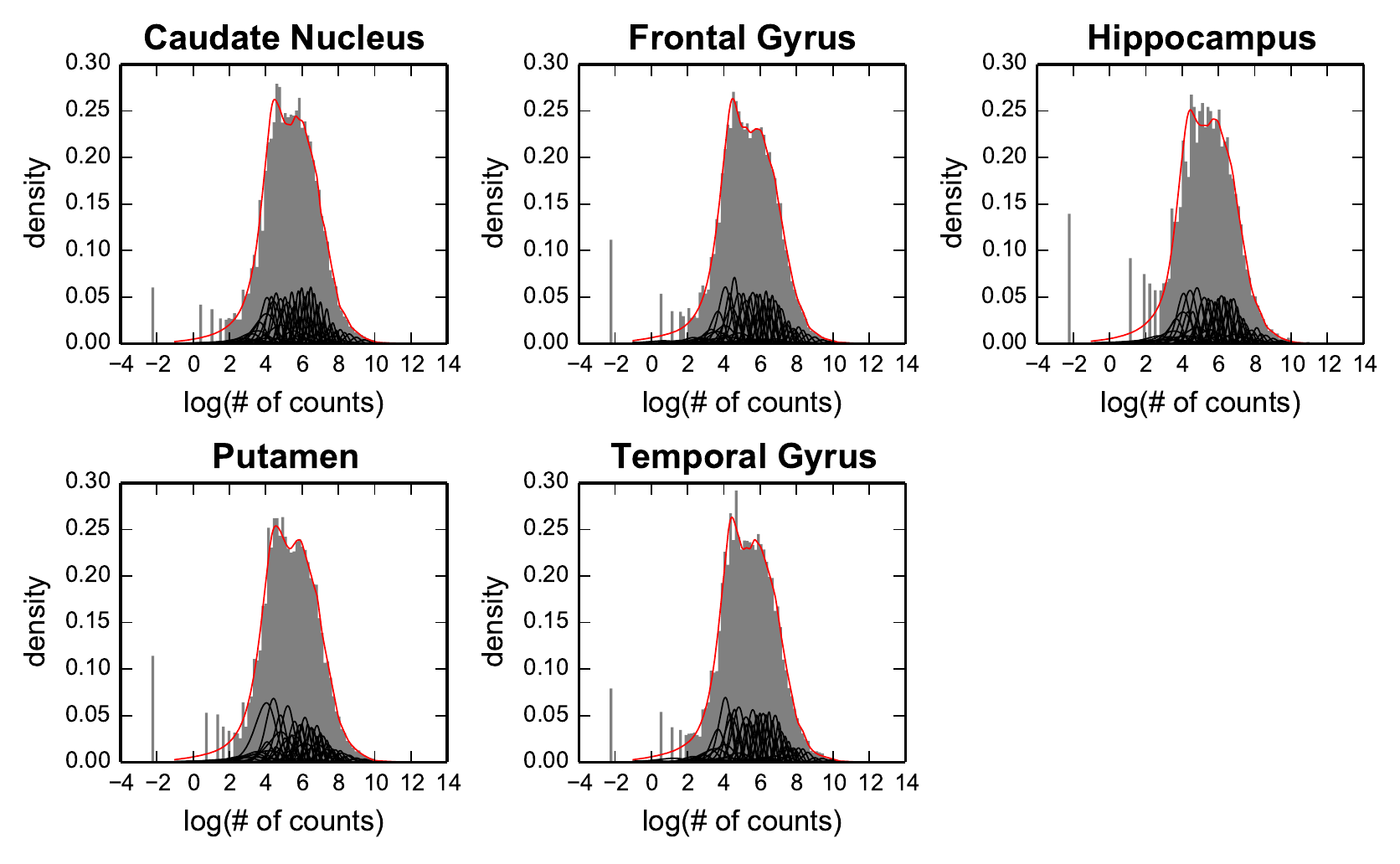} \\
	\textbf{Figure 5}
\end{center}	
\end{figure}

\begin{figure}[h]
\begin{center}
	\includegraphics{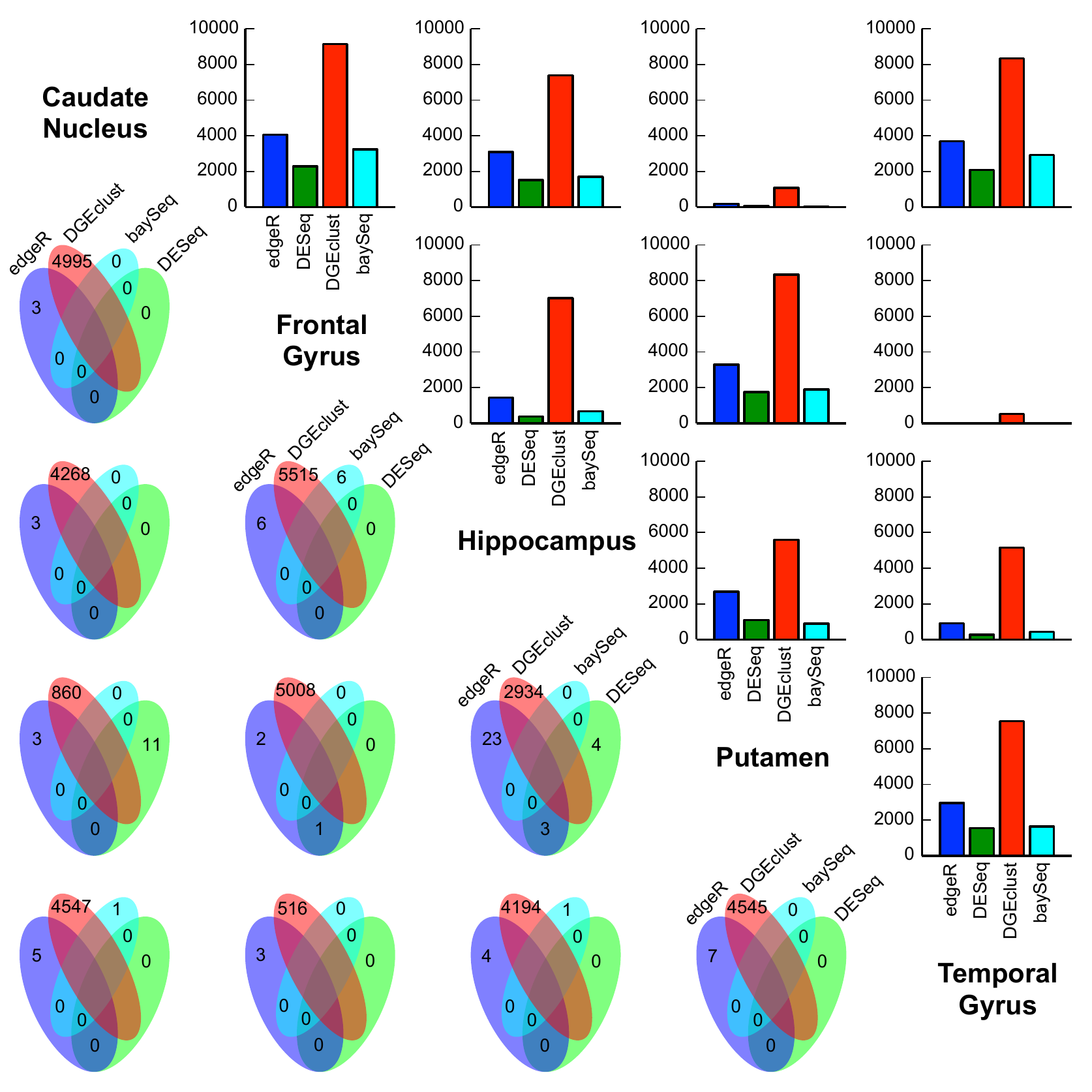} \\
	\textbf{Figure 6}
\end{center}	
\end{figure}

\begin{figure}[h]
\begin{center}
	\includegraphics{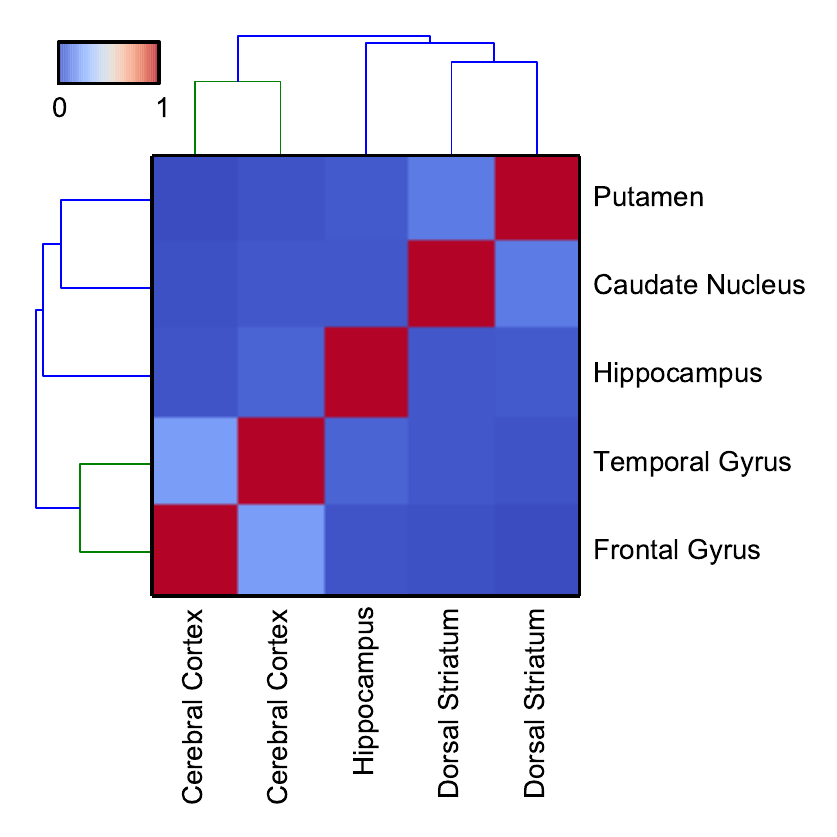} \\
	\textbf{Figure 7}
\end{center}	
\end{figure}

\end{bmcformat}
\end{document}